\documentclass[aps,prd,twocolumn,superscriptaddress]{revtex4}
\usepackage{amsmath,epsfig,mathrsfs,graphicx,amssymb,slashed}

\def\be#1\ee{\begin{equation}#1\end{equation}}

\newcommand{\ba}{\begin{eqnarray} }
\newcommand{\ea}{\end{eqnarray} }

\begin{document}
\title{Local bosonization of massive fermions in three spatial dimensions with rotation invariance}
\author{Adam Bednorz}
\email{Adam.Bednorz@fuw.edu.pl}
\affiliation{Faculty of Physics, University of Warsaw
ul. Pasteura 5, PL02-093 Warsaw, Poland}

\date{\today}

\begin{abstract}
In relativistic quantum field theory particles of half-integer spin must obey Fermi-Dirac statistics. Their quantum operators must anticommute at spacelike separation in contrast to commuting physical observables. We show that Fermi-Dirac spin $1/2$ operators can be emergent in a fully commuting field theory forming directed strings and loops of spin 0 and 1 constituents, reproducing massive Dirac dynamics with background fields. Such underlying description may violate relativistic invariance but there are no manifest interactions at a distance and rotation symmetry remains preserved. We show that under some constraints on the model there exists a well-defined ground state -- Fermi sea that it is stable -- fermions cannot convert to bosons.
\end{abstract}
\maketitle

\section{Introduction}
\label{sec:intro}

Fully relativistic quantum field theories, such as electrodynamics imply the existence of two kinds of fields: commuting bosons (Bose-Einstein operators) and anticommuting fermions (Fermi-Dirac operators). The former are realized for integer spins while the latter for half-integer. The proof of spin-statistics correspondence requires relativistic invariance and energy positivity \cite{qft1,qft2,qft3,qft4,qft5} while relativity is a postulate imposed on quantum field theories \cite{wight}. However, physically observable quantities correspond only to commuting operators so fermion operators are solely elements of mathematical descriptions -- they are not directly observable (unless one takes two fermion operators forming usually a nonlocal object). The division into fermions and bosons remains in all modern theories, including standard model, string or superstring and $M$-theory \cite{string1,string2,string3,string4}.

Some time ago it has been proposed a theory reducing fermions to composite states of bosons -- string-nets -- at very high energy/momentum scale \cite{wen1,wen2}. The rough idea is that the fermions are emergent as endpoints of strings fluctuating in empty space. Even sacrificing relativity this concept is an interesting alternative to standard string theories, where fermions are always fundamental -- not emergent (even if supported by spin-statistics theorem and supersymmetry). Although the idea is an attractive alternative direction of progress in quantum field theory including quantum gravity \cite{kon1,kon2}, the so far developed models (mostly in $2$ spatial dimensions, usually on lattice) fail to address clearly many important issues:
\begin{itemize}
\item symmetry (relativity, rotation in 3D)
\item emergence of the spin $1/2$ out of spin $0$ and $1$ constituents and antisymmetry
\item recovering effective massive fermions 
\item depth of the Fermi sea
\item collapse of fermions to bosons
\item background field
\end{itemize}

In this paper, we will construct a general family of models, addressing these points, identifying the parameter range of validity.
A general property of the models presented here is lack of full relativistic invariance. It is known that relativity considerably reduces available composite theories \cite{wewi}. However, this cannot invalidate our models because
the models are still local in the sense of lack of action at a distance and some further improvements like extra dimensions may restore full invariance.
The locality means here that the Hamiltonian connects configurations differing only in a finite range (i.e. $\hat{H}=\int d\vec{r}\mathcal H(\vec{r})$
where $\mathcal H(\vec{r})$ depends only on the part of the configuration in a generally bounded distance from $\vec{r}$).
We will work in $3$ spatial and $1$ temporal dimension. Instead of action and path integrals \cite{ab04}, being often the starting point for usual strings, our whole model is Hamiltonian-based. As in earlier works, the basic object remains a directed string but we will show a correct construction of a Hamiltonian which preserves rotation symmetry $SO(3)$ and recovers effective low-energy Dirac dynamics. The spins $1/2$ at the string endpoints combine through
spinless singlet states along the string to integer-spin structures, forming an $SU(2)$ Wilson line/loop \cite{wilson}. Therefore
the only constituents are here integer spin bosons. The Hamiltonian couples locally different strings by a kind of small sheet/plaquette terms \cite{kogut}, remaining $SO(3)$-invariant. Special terms of the Hamiltonian form the bottom of the Fermi sea and prevent from transition into a bosonic state, i.e. collapse to the symmetric state of lower energy. Incorporation of background potentials allows to replace them with fields.
 The model is mainly tailored to electrodynamics but its key features make it possible to generalize them to other theories. We failed to present a Lorentz invariant model but we cannot judge if such construction is just more complicated or impossible.

The paper is organized as follows. First, the standard description of fermions in quantum electrodynamics is recalled.
Then, the model of directed strings is proposed and the goal -- one-to-one correspondence between integer spin bosonic states in the string-net and spin-$1/2$ fermions is stated.
The necessary terms of the Hamiltonian are outlined in the next sections with  technical details left in Appendix. Finally, we reconstruct effective Dirac Hamiltonian, including background electromagnetic fields. We close the paper with the discussion of the high-energy deviations and proposed further development of the models.

\section{Fermions in quantum electrodynamics}

The standard theory of free fermions (e.g. electrons and positrons) of mass $m$ starts with Dirac wave equation
\begin{equation}
(\gamma^\mu(i\partial_\mu-A_\mu)-m)\psi(x)=0\label{dirac}
\end{equation}
where $\psi$ is a four-component field in spacetime defined as $x=(x^0=ct,x^1,x^2,x^3)$
with $\vec{x}=(x^1,x^2,x^3)$ representing spatial position while $x^0$ is time $t$ multiplied by the speed of light $c=1$ ($x$ can be replaced by $y$ or $r$).
Here we use standard conventions, including flat metric tensor $g^{\mu\nu}=g_{\mu\nu}=\mathrm{diag}(1,-1,-1,-1)$, summation convention
$X^\mu Y_\mu=\sum_\mu X^\mu Y_\mu$, derivatives $\partial_\mu=\partial/\partial x^\mu$, four-potential $A_\mu(x)$ (with charge included) and Dirac $4\times 4$ matrices $\gamma^\mu$ (Hermitian $\gamma^0$ and anti-Hermitian $\gamma^{1,2,3}$) satisfying anticommutation rule $\{\gamma^\mu,\gamma^\nu\}=2g^{\mu\nu}$. Here we adopt Weyl convention
\begin{equation}
\gamma^0=
\begin{pmatrix}
0&I\\
I&0\end{pmatrix},\: \gamma^{i}=\begin{pmatrix}
0&\sigma^i\\
-\sigma^i&0\end{pmatrix}
\end{equation}
with Pauli matrices
\begin{eqnarray}
&&I=\begin{pmatrix}
1&0\\
0&1\end{pmatrix},\:\sigma^1=\begin{pmatrix}
0&1\\
1&0\end{pmatrix},
\nonumber\\
&&\sigma^2=\begin{pmatrix}
0&-i\\
i&0\end{pmatrix},\:\sigma^3=\begin{pmatrix}
1&0\\
0&-1\end{pmatrix}
\end{eqnarray}
We distinguish left/right two-dimensional components, $\psi_{L/R}$, respectively, in $\psi=(\psi_L,\psi_R)^T$.

The problem of anticommutation appears at the level of second quantization.
One constructs Lagrangian density in the form
\begin{equation}
\mathcal L(x)=\bar{\psi}(x)(\gamma^\mu(i\partial_\mu-A_\mu)-m)\psi(x)
\end{equation}
with $\bar{\psi}=\psi^\dag\gamma^0$ and Hamiltonian \cite{qft1,qft2,qft3,qft4,qft5}
\begin{equation}
\hat{H}(x^0)=\int d\vec{x} \hat{\bar{\psi}}(\vec{x})(\vec{\gamma}\cdot(-\vec{A}(x)-i\nabla)+m+\gamma_0A_0(x))\hat{\psi}(\vec{x})\label{ham}
\end{equation}
Here we work in $3D$ spatial space, $\hbar=1$, $d\vec{x}=dx^1dx^2dx^3$, $\vec{X}=(X^1,X^2,X^3)$, $\nabla=(\partial_1,\partial_2,\partial_3)$ with the standard scalar product
$\vec{X}\cdot\vec{Y}=\sum_{i=1,2,3}X^iY^i$. The standard spin-statistics theorem, which assumes relativity and positive energy,  implies anticommutation rule
\begin{equation}
\{\hat{\psi}^\dag_a(\vec{x}),\hat{\psi}_b(\vec{y})\}=\delta_{ab}\delta(\vec{x}-\vec{y}),\:
\{\hat{\psi}_a(\vec{x}),\hat{\psi}_b(\vec{y})\}=0\label{ant}
\end{equation}
or, using equivalent path integral formulation 
\begin{equation}
\int D\psi \exp\int i\mathcal L(x) d^4x
\end{equation}
the integration runs over Grassmann variables $\psi_a(x)\psi_b(y)=-\psi_b(y)\psi_a(x)$ and $dx=dx^0 d\vec{x}$.
The dynamics under (\ref{ham}) is usually described by diagonalization of $\hat{\psi}\to\hat{\psi}_\epsilon$ using eigenstates of single-particle Dirac equation (\ref{dirac})
with $i\partial_t=\epsilon$ as single-particle energy. The ground state has all states with $\epsilon<0$ occupied while
all other states can be written using anticommuting eigenstate operators $\hat{\psi}_\epsilon$. For time-dependent potentials $A$ one uses time-dependent
orthonormal solutions of (\ref{dirac}) \cite{qft1,qft2,qft3,qft4,qft5}.

The aim of this paper is to construct a model of directed strings, whose dynamics at low energies reduces effectively to Dirac Hamiltonian (\ref{ham}) with anticommutation rules (\ref{ant}). Our model will not be relativistic so the standard spin-statistics theorem does not apply and so the anticommutation must be justified in a different way.

\section{Directed string}

\begin{figure}
\includegraphics[scale=.5]{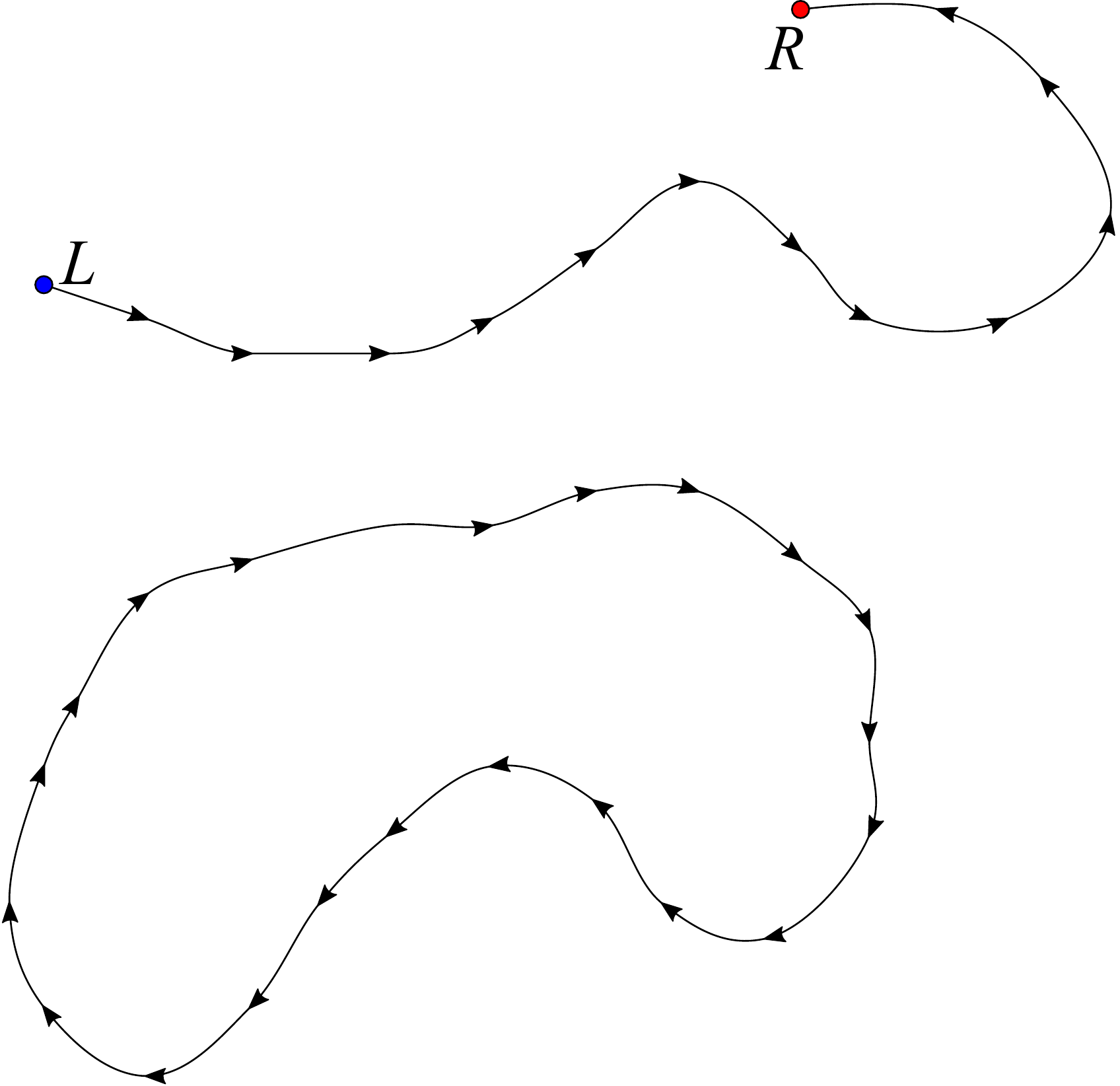}
\caption{General idea of directed string. The open string has left and right opposite particles at its ends.
The closed string has its direction with no distinguished endpoint}
\label{st0}
\end{figure}

As left and right $\psi_{R/L}$, we define left/right-handed operators $\hat{\psi}_{L/R}$  and $\hat{\psi}^\dag_{L/R}$ in 
\begin{equation}
\hat{\psi}=\begin{pmatrix}\hat{\psi}_L\\
\hat{\psi}_R\end{pmatrix},\: \hat{\psi}^\dag=(\hat{\psi}^\dag_L,\hat{\psi}^\dag_R).
\end{equation}
Let us start with some initial state (not necessarily ground) with all right-handed states empty and all left-handed states occupied. The state is $|\Omega\rangle$
with the property 
\begin{equation}
\hat{\psi}_R|\Omega\rangle=\hat{\psi}^\dag_L|\Omega\rangle=0
\end{equation}
Now the basic excitation reads $\hat{\psi}_{La}(\vec{x})\hat{\psi}^\dag_{Rb}(\vec{y})$ or 
\begin{equation}
|\vec{x}_{La}\vec{y}_{Rb}\rangle=\hat{\psi}_{La}(\vec{x})\hat{\psi}^\dag_{Rb}(\vec{y})|\Omega\rangle
\end{equation} where $a$ and $b$ are indices in the $2$-dimensional respective spin space. Let us identify this excitation with a string directed from $\vec{x}$ (left point) to $\vec{y}$ (right point). Taking just a straight line
would suffice but then locality is manifestly broken. It will be anyway broken anyway in the relativistic sense but we will assume finite range of the Hamiltonian. Therefore we consider the whole family of continuous directed strings between these points. 
Such strings are homotopic to an interval so they are open.
We allow additional separate directed closed strings (loops), see Fig. \ref{st0}. We do not yet impose any condition on the shape of the strings and the number of loops but such constraint will appear in particular models discussed later. 

\begin{figure}
\includegraphics[scale=.5]{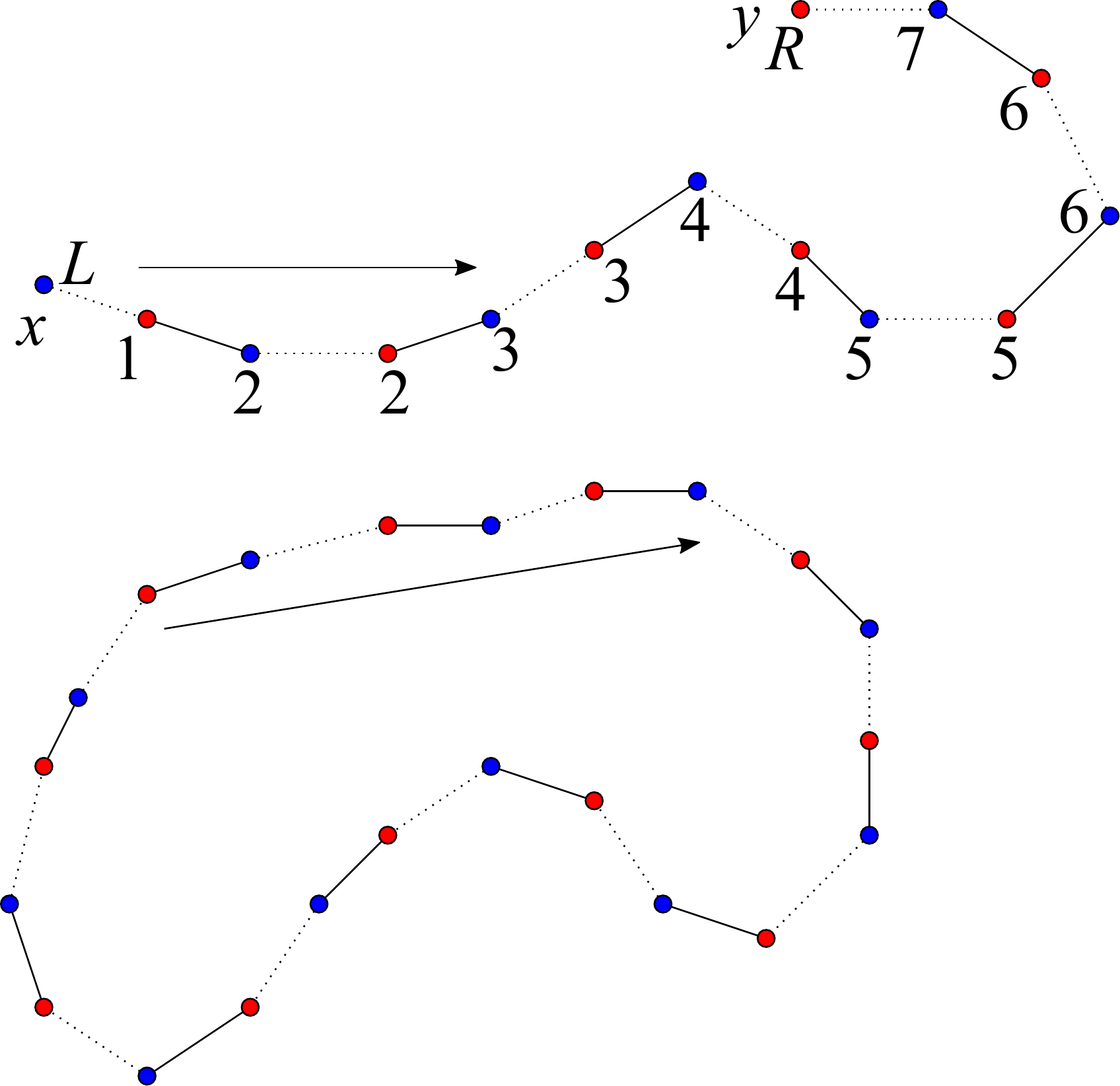}
\caption{The string with inserted singlets (solid lines) both in the case of an open and closed string.}
\label{st1}
\end{figure}

\begin{figure}
\includegraphics[scale=.5]{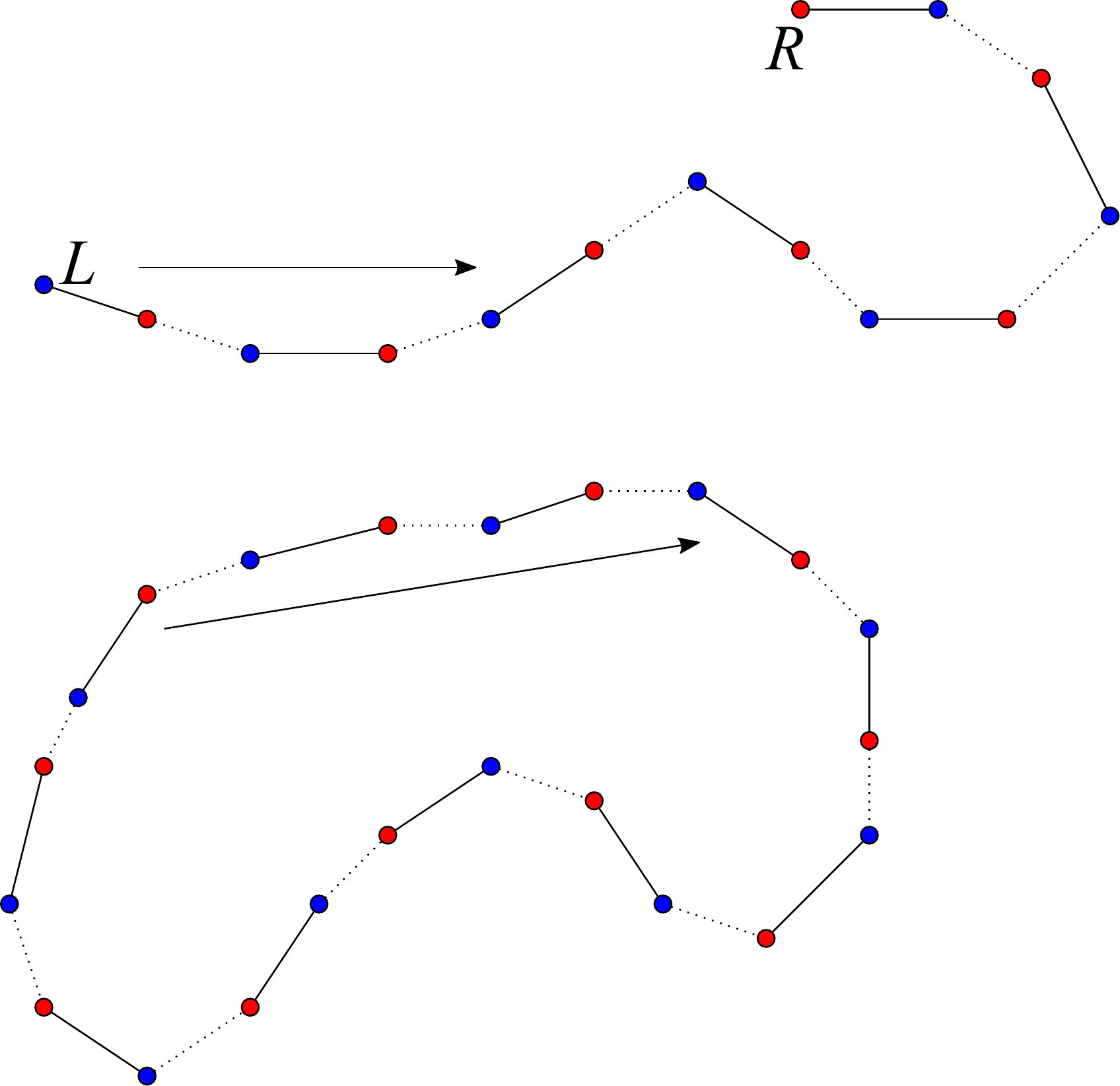}
\caption{The states from Fig. \ref{st1} with swapped links to combine the endpoints into singlets and triplets.}
\label{st2}
\end{figure}

Our aim is to find a Hamiltonian model of the strings that leads to the effective Dirac dynamics in low energy approximation.
The basic element of such a model will be the directed string with spin $1/2$ ends (if open). In particular the string will be temporarily represented
by $2n$ local spins $1/2$ interchanging between $R$ and $L$, i.e. a generic state reads
\begin{equation}
|\psi\{\vec{r}\}\rangle=\prod_{j=1}^n|a_j\rangle_L|b_j\rangle_R
\end{equation}
with the string going through a chain of $n$ points $\vec{x}=\vec{r}_1,\vec{r}_2,\dots,\vec{r}_n=\vec{y}$ such that subsequent points are close to each other (in the case of a closed loop $\vec{r}_1$ follows $\vec{r}_n$)
and $a_j,b_j=\pm$, corresponding to states $|a\rangle_L$ and $| b\rangle_R$ with $a,b=\pm$ (basis order $|+\rangle,|-\rangle$) in the above mentioned excitation. For a moment the chain is finite but we will consider a continuum limit. There are in principle $2^n$ possible states for a given string trajectory $\vec{r}$. We want to reduce the degeneracy to a $2\times 2$ combination of endpoint states $|a_1\rangle_L$ and $|b_n\rangle_R$. Such states can be obtained by combining the intermediate state into singlets $(|+_{Rj}+_{Lj+1}\rangle+|-_{Rj}-_{Lj+1}\rangle)/\sqrt{2}$, splitting $\vec{r}_j\to\vec{r}_{Lj},\vec{r}_{Rj}$ with $\vec{x}=\vec{r}_{L1}$ and $\vec{y}=\vec{r}_{Rn}$, see Fig. \ref{st1}. This definition of the singlet differs from familiar $|+-\rangle-|-+\rangle$ because of the transpose used in the spinor convention here. This state reads
\begin{equation}
|\psi\{\vec{r}\}\rangle_{ab}=\frac{|a\rangle_L|b\rangle_R}
 {\sqrt{2}^{n-1}}\prod_{j=1}^{n-1}(|+_{Rj}+_{Lj+1}\rangle+|-_{Rj}-_{Lj+1}\rangle)\label{ope}
\end{equation}
For a closed string (denoted by subscript $\ell$) we have only singlets
\begin{equation}
|\psi\{\vec{r}\}\rangle_\ell= 2^{-n/2}\prod_{j=1}^{n}(|+_{Rj}-_{Lj+1}\rangle+|-_{Rj}+_{Lj+1}\rangle)\label{clo}
\end{equation}
with $n\equiv 0$. Now we apply \emph{spin swapping}, see Fig. \ref{st2}, i.e. couple the pairs of the same $j$ and project onto one of the singlet and triplet states
\begin{eqnarray}
&&\sqrt{2}|0_j\rangle=|+_{Rj}+_{Lj}\rangle+|-_{Rj}-_{Lj}\rangle,\nonumber\\
&&\sqrt{2}|3_j\rangle=|+_{Rj}+_{Lj}\rangle-|-_{Rj}-_{Lj}\rangle,\nonumber\\
&&\sqrt{2}|1_j\rangle=|+_{Rj}-_{Lj}\rangle+|-_{Rj}+_{Lj}\rangle,\\
&&\sqrt{2}|2_j\rangle=i|-_{Rj}+_{Lj}\rangle-i|+_{Rj}-_{Lj}\rangle,\nonumber
\end{eqnarray}
The generic state in the space of these states along the string will be denoted
\begin{equation}
|c\rangle=\prod_{j=1}^n |c_j\rangle
\end{equation}
with $c=0,1,2,3$. Now the states (\ref{ope}) and (\ref{clo}) can be expressed as entangled states of singlets and triplets with
\begin{eqnarray}
&&\langle c|\psi\rangle_{ab}=2^{1/2-n}(\sigma_{c_1}\sigma_{c_2}\cdots\sigma_{c_n})_{ab}\nonumber\\
&&\langle c|\psi\rangle=2^{-n}\mathrm{Tr}\sigma_{c_1}\sigma_{c_2}\cdots\sigma_{c_n}
\end{eqnarray}
ignoring poistion $\vec{r}$ for a moment.
Singlet and triplet states correspond to the total integer spin, $0$ and $1$, respectively, and so they belong already to the bosonic description.
It will be clear later when reconstructing Dirac equation.
From now on, the singlets and triplets become the bosonic constituent systems of the whole dynamics. There are neither
fundamental spin $1/2$ particles nor antisymmetric (fermionic) states. Spin $1/2$ antisymmetric fermions will \emph{emerge} effectively at low energy
as collective states of integer-spin bosons.
This description can be generalized further, assuming almost arbitrary space along the string, where we can define a complex scalar $v^0$ and a vector
$\vec{v}=(v^1,v^2,v^3)$ to decompose
\begin{eqnarray}
&&\langle v|\psi\rangle_{ab}\propto\langle a|V_1V_2\cdots V_n|b\rangle\nonumber\\
&&\langle v|\psi\rangle_\ell\propto\mathrm{Tr}V_1V_2\cdots V_n
\end{eqnarray}
with $V_j=\sum_k v^k_j\sigma_k\neq 0$, being a general nonzero complex $2\times 2$ matrix. The string and the sequence of $V$ matrices can be defined continuously, with the string parametrized by real $s$ on an interval. Then the string position is $\vec{r}(s)$ while $V_j\to I+i\vec{\sigma}\cdot\vec{v}ds$ (removing $c$-number $v_0$) with the complex vector function $\vec{v}(s)$. Then we get $SU(2)$ Wilson line/loop \cite{wilson} matrix
\begin{equation}
V_1V_2\cdots V_n\to V=\mathcal P\exp\int ids \vec{\sigma}\cdot\vec{v}(s)
\end{equation}
where $\mathcal P$ denotes ordering along growing $s$ in the power expansion, i.e. $\vec{\sigma}\cdot\vec{v}(s)\vec{\sigma}\cdot\vec{v}(s')$ for $s'>s$.
We will define $|\psi\rangle_{ab}$ as the collective state of strings  using the above representation as building blocks.

\section{Collective states of strings}

The example with spin swapping shows that the effective spinor state (particular values of $a$ and $b$ at the endpoints)
can be an entangled state of the states defined by $v$ and specific trajectories $r$. Suppose the space of allowed $v$ is given.
In the case of a chain it can be discrete, e.g. $v^{(0)}=(1,0,0,0)$, $v^{(1)}=(0,1,0,0)$, $v^{(2)}=(0,0,1,0)$, $v^{(3)}=(0,0,0,1)$, or continuous, e.g. $v=(1,\vec{v})$
with real unit $\vec{v}$ or completely arbitrary complex four-vector $v$. In the continuous string $\vec{v}$ can be real or otherwise restricted. The configuration space contains string position $r$ and matrix function $v$ which
is a chain of points $\vec{r}_1,...,\vec{r}_n$ and matrices $v_1,...,v_n$ or functions $\vec{r}(s)$ and $\vec{v}(s)$ with $s$ being 1-dimensional real parameter along the string between for $s\in [s_L,s_R]$. The endpoints are $\vec{x}\equiv \vec{r}_L=\vec{r}(s_L)$ and $\vec{y}\equiv\vec{r}_R=\vec{r}(s_R)$,
In the case of the loop $\vec{r}_L=\vec{r}_R=\vec{r}(s_L)=\vec{r}(s_R)$ with $s\equiv s+s_R-s_L$ (loop topology).
The configuration state $|rv\rangle$ denotes vector functions $\vec{r}(s)$ (real) and $\vec{v}(s)$ (real, imaginary, complex or otherwise restricted)
for all available $s$,
orthonormal in the functional sense $\langle r'v'|rv\rangle=\delta(\vec{r}'-\vec{r})\delta(\vec{v}'-\vec{v})$
in the functional measure $\int DrDv \delta(\vec{r})\delta(\vec{v})=1$ with completeness $\hat{1}=\int DrDv |rv\rangle\langle rv|$.
We shall \emph{assume} the effective spinor state and loop state of the form
\begin{eqnarray}
&&\langle rv|\vec{x}_{La}\vec{y}_{Rb}\rangle= f(r,v)V_{ab}\nonumber\\
&&\langle rv|\Omega\rangle= f(r,v)\mathrm{Tr}V
\end{eqnarray}
with $V=V_1V_2\cdots V_n$ in the case of a chain and $V=\mathcal P\exp\int ids\vec{\sigma}\cdot\vec{v}(s)$ in the continuous case.
Here $f(r,v)$ is an assumed wave function, quite general with only several reasonable conditions
\begin{itemize}
\item $\vec{r}_{L1}=\vec{x}$, $\vec{r}_{Rn}=\vec{y}$ in an open string, $0\equiv n$ in a closed loop
\item normalization (decay at large values), e.g. $-\int ds |\vec{v}(s)|^2$ term in $\ln f$
\item rotation invariance, i.e. $f$ must be a scalar function of $r$ and $v$
\item translation invariance, i.e. $f(\vec{r},\vec{v})=f(\vec{r}+\vec{r}_0,\vec{v})$
for an arbitrary constant vector $\vec{r}_0$.
\end{itemize}

Of course, the collective states can contain a single open string \emph{and} an arbitrary number of closed loops. 
In general we can a have an arbitrary number of fundamental excitations, i.e.
\begin{eqnarray}
&&|\vec{x}_{1},a_1,\vec{x}_{2},a_2\dots,\vec{x}_N, a_N;\vec{y}_1,b_1,\vec{y}_2,b_2,\dots,\vec{y}_N,b_N\rangle\nonumber\\
&&
=\hat{\psi}_{La_1}(\vec{x}_1)\hat{\psi}^\dag_{Rb_1}(\vec{y}_1)\cdots \hat{\psi}_{La_N}(\vec{x}_N)\hat{\psi}^\dag_{Rb_N}(\vec{y}_N)|\Omega\rangle
\label{multi}
\end{eqnarray}
Due to Fermion anticommutation rule we have the Pauli property
\begin{eqnarray}
&&|\vec{x}_{1},a_1,\dots,\vec{x}_N, a_N;\vec{y}_{\tau(1)},b_{\tau(1)},\dots,\vec{y}_{\tau(N)},b_{\tau(N)}\rangle
\nonumber\\=
&&\mathrm{sgn} \tau|\vec{x}_{1},a_1,\dots,\vec{x}_N, a_N;\vec{y}_1,b_1,\dots,\vec{y}_N,b_N\rangle
\end{eqnarray}
for the permutation $\tau$. 
We will assume that the state (\ref{multi}) is a collective state of $N$ open strings and an arbitrary number of closed strings.
The reference empty state $|\Omega\rangle$ is represented by only closed loops.
Each open string starts at some $\vec{x}_j$ and ends at $\vec{y}_{\tau(j)}$ with some permutation $\tau$.
Then the collective state (\ref{multi}) reads
\begin{eqnarray}
&&\sum_{r,v,S,\ell}(-1)^M\mathrm{sgn}\tau f(r,v)|rv\rangle\times\label{stri}\\
&&(V_{S_1})_{a_1b_{\tau(1)}}\cdots (V_{S_N})_{a_Nb_{\tau(N)}}
\mathrm{Tr}V_{\ell_1}\cdots \mathrm{Tr}V_{\ell_M}\nonumber\nonumber
\end{eqnarray}
with open strings $S_j$ from $\vec{x}_j$ to $\vec{y}_{\tau(j)}$ and closed loops $\ell_j$ and \emph{local and rotationally invariant}
function $f$. It means in general that $f$ must be normalizable (i.e. $\int DrDv |f|^2<\infty$) and have \emph{cluster property}, i.e. it is a product of local functions,
involving $v$ for which $\vec{r}$ are  close. In particular
\begin{eqnarray}
&&\ln f(r,v)=\int ds \kappa_1(\vec{r}(s),\vec{v}(s),s)+ \\
&&\int dsds' \kappa_2(\vec{r}(s),\vec{r}(s'),\vec{v}(s),\vec{v}(s'),s,s')+...\nonumber
\end{eqnarray}
with $\kappa_2$ vanishing at large $\vec{r}(s)-\vec{r}(s')$ or $|s-s'|$.
Some reasonable terms that can appear in $-\ln f$ are
\begin{equation}
\int ds (\alpha|d\vec{r}(s)/ds|^2+\beta|d\vec{v}(s)/ds|^2+\eta|\vec{v}(s)|^2)
 \end{equation}
This condition is essential to achieve locality. Otherwise, we could apply just nonlocal coupling of pairs of particles and claim bosonization.
Instead, we want to show that the underlying model is formally local in space (but not necessarily in the relativistic sense of invariance and communication limited by the speed of light). The \emph{nominal length} of the continuous string is $s_R-s_L$ although the actual length $\int ds|d\vec{r}/ds|$ may be different
(the string can be stretched or squeezed).

In our model $f$ will be a product of individual strings/loops i.e.
\begin{eqnarray}
&&f(r,v)= \mathcal Z f(r_{S_1},v_{S_1}) \cdots f(r_{S_N},v_{S_N})\times\nonumber\\
&& f(r_{\ell_1},v_{\ell_1}) \cdots f(r_{\ell_N},v_{\ell_N})
\end{eqnarray}
with the normalization factor $\mathcal Z$
but one can also include factors modifying $f$ when strings are close to each other at some point.

The states are defined in a $3D$ box of dimensions $\Omega_1$, $\Omega_2$, $\Omega_3$,  with periodic boundary conditions $\vec{r}+(n_1\Omega_1,n_2\Omega_2,n_3\Omega_3)\equiv\vec{r}$
for arbitrary integers $n_1,n_2,n_3$. In the thermodynamic limit $\Omega_1,\Omega_2,\Omega_3\to\infty$ we keep $|S|\propto \Omega=\Omega_1\Omega_2\Omega_3$, where $|S|$ is the total
nominal length of all loops and strings (counted along parameter $s$).

In the construction of string states, it is important that they are not just a bunch of vectorlike particles scattered in space but they contain information about string \emph{order}. In other words, every segment of the string contains also information about its successor and predecessor in a chain
or direction of a continuous curve.

\section{Basic Hamiltonian}

The existence states constructed in the previous section must follow from the structure of a model Hamiltonian.
The general form is
\begin{equation}
\hat{H}=\int Dr'Dv'DrDv h(\vec{r}'\vec{v}';\vec{r}\vec{v})|r'v'\rangle\langle rv| \label{kerne}
\end{equation}
with local, rotationally invariant kernel function $h$.
Optionally, functional derivatives like $\delta_{\vec{r}}=\delta/\delta\vec{r}$ acting or either $|rv\rangle$ or $\langle rv|$ are allowed.
We will construct such a Hamiltonian $\hat{H}$
that all the states (\ref{stri}) are annihilated by $\hat{H}$ (i.e. they are eigenstates with eigenvalue $0$), while all other states have strictly positive
eigenvalues, larger than the energy scale of the effective theory. We also stress that the family of Hamiltonian reproducing the low-energy
collective states is quite large, analogously to quantum phase transitions, and the model presented here is only one example yet with many freedom parameters.

Before the proper construction let us outline its idea in the simple example -- harmonic oscillator.
The ground wave function of $1$-dimensional oscillator has the form $e^{-\alpha x^2}$. Applying derivative (local) operator $d/dx$ we obtain $-2\alpha xe^{-\alpha x^2}$
so it is obvious that $\hat{c}=d/dx+2\alpha x$ annihilates the state. Now $\hat{H}=\hat{c}^\dag\hat{c}$ is a positive operator and its only $0$-eigenvalue
eigenstates $\psi(x)$ must satisfy  $\hat{c}\psi=0$ which gives back the assumed state as the only solution. The other eigenvalues must be nonzero.
In the case of the oscillator we are able to find them exactly but in general it is possible to make an estimate. Note that those positive eigenvalues can be scaled up arbitrarily multiplying $\hat{H}$ by an appropriate factor. A multidimensional oscillator ground state $e^{-\alpha|\vec{x}|^2}$ is distinguished by
defining $\hat{\vec{c}}=(\nabla_x+2\alpha\vec{x})$ and $\hat{H}=\hat{\vec{c}}^\dag\cdot\hat{\vec{c}}$ so the idea easily extends to an arbitrary state and space. 

\begin{figure}
\includegraphics[scale=.4]{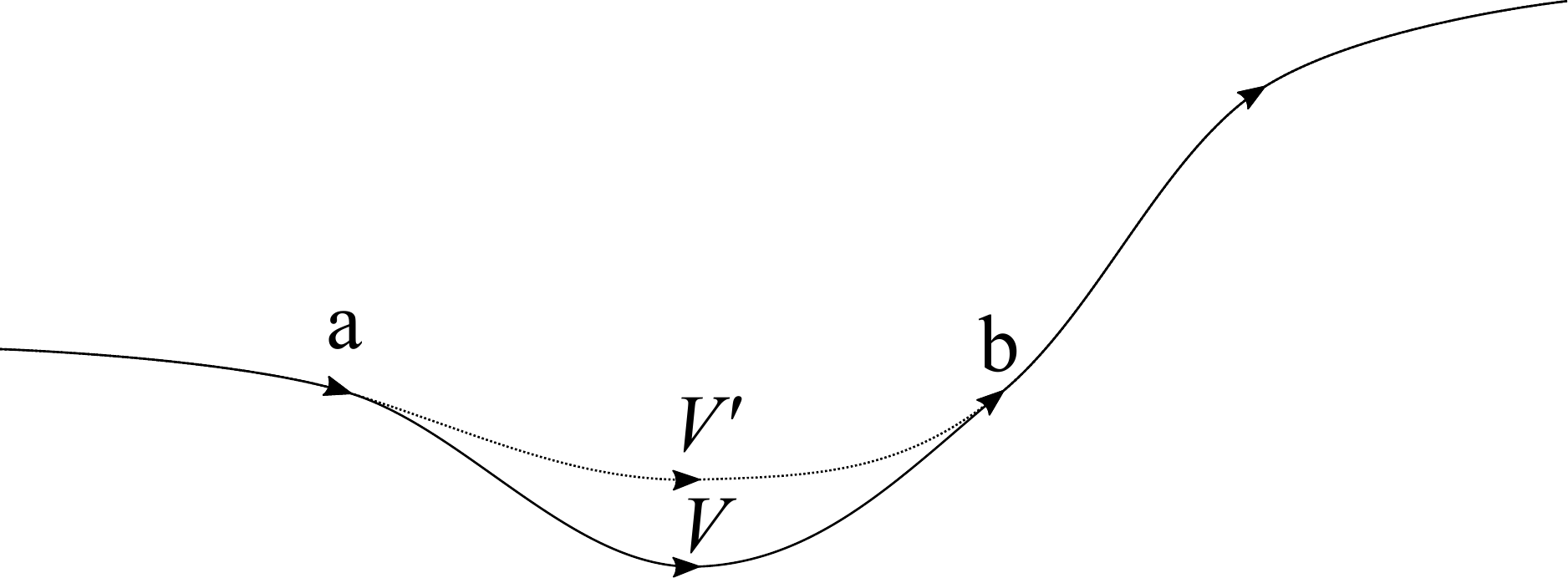}
\caption{Local wiggling of a string fragment (solid $\to$ dotted).}
\label{wi1}
\end{figure}

\begin{figure}
\includegraphics[scale=.4]{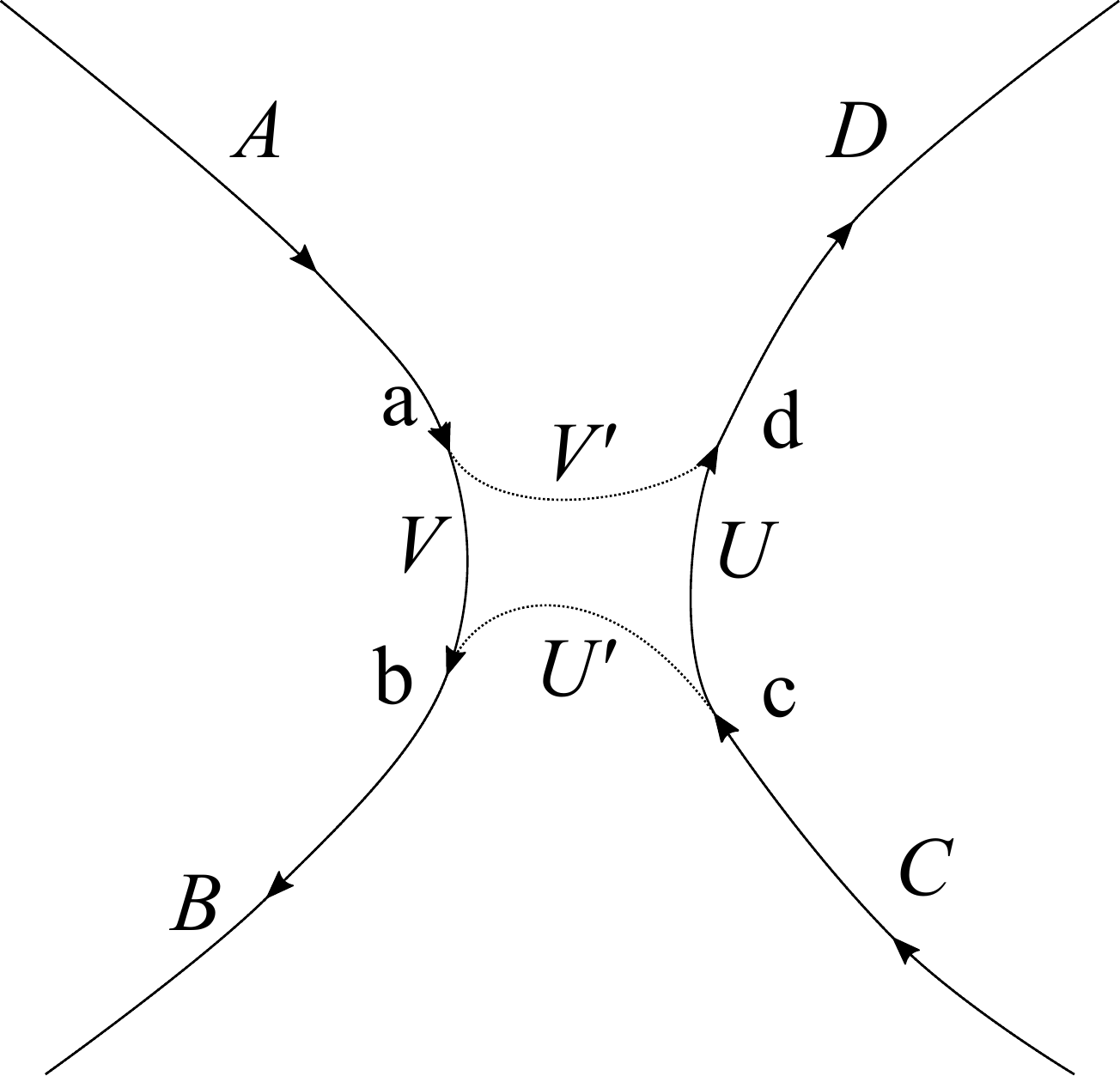}
\caption{Local swap between fragments of two different strings (solid $\to$ dotted).}
\label{wi2}
\end{figure}

We shall apply that above outlined construction to the family of states (\ref{stri}). Just like in the harmonic oscillator, we have to find  \emph{local}
operators connecting different constituent states, e.g. with a string (or a couple of them) wiggled (or swapped) inside a localized volume, see Figs. \ref{wi1}
and \ref{wi2}. Wiggling means combining parts of the string sequence (link) $\bar V=V_lV_{l+1}\cdots V_m$ and $\bar V'=V'_{l'}V'_{l'+1}\cdots V'_{m'}$
or $\bar V=\mathcal P\exp\int_w ids\vec{\sigma}\cdot\vec{v}$ to $\bar V'=\mathcal P\exp\int_w ids\vec{\sigma}\cdot\vec{v}'$ (subscript $w$ indicated restriction to the wiggled part) and corresponding parts wave functions $\bar f$ and $\bar f'$  depending only on the string part around the wiggled part while leaving the rest unchanged, i.e. $f=g\bar f$ and $f'=g\bar f'$ with $g$ factor covering the not wiggled rest of the string(s).
Both $\bar{f}$ and $\bar f'$ must depend only on the local neighborhood of the wiggled part. In this case, the nominal length remains constant.
More generally we will consider a family $\bar f^{1}$, $\bar V^{1}$, $\bar f^{2}$ $\bar V^{2}$,...,$\bar f^{K}$ $\bar V^{K}$ with $K>1$
and $f^{j}=g\bar{f}^{j}$. For $K=2$ we can assign $\bar f^{1}=\bar f$, $\bar V^{1}=\bar V$ and  $\bar f^2=\bar f'$,$\bar V^{(2)}=\bar V'$.
Each matrix $V^{j}$ is $2\times 2$ dimensional.
Let us consider Slater determinant in $(2\times 2)^K$ dimensional space \cite{slater}
\begin{equation}
\mathcal W_K=\sum_\sigma\mathrm{sgn}\tau \bar f^{\tau(1)}\bar V^{\tau(1)}\bar f^{\tau(2)}\bar V^{\tau(2)}\cdots \bar f^{\tau(K)}\bar V^{\tau(K)}\label{slat}
\end{equation}
with the sum over permutations $\tau$.
It is clear that the determinant is zero for $K>4$ because there are maximally $4$ independent $2\times 2$ matrices.

Let us define annihilation operator acting on $2\times 2$ matrices with a $(2\times 2)^{K-1}$ matrix as a result
\begin{equation}
\hat{c}(r,v)=\sum_\tau\mathrm{sgn}\tau \bar f^{\tau(2)}\bar V^{\tau(2)}\cdots \bar f^{\tau(K)}\bar V^{\tau(K)} \langle rv^{\tau(1)}|\label{slat2}
\end{equation}
It is clear that it annihilates the postulated ground states for $K>4$ but $\hat{c}$ is zero identically for $K>5$ so the best choice is $K=5$.
For $K<5$ we have to add the condition that $\mathcal W_K=0$ by e.g. $\delta(\mathcal W_K)$ modeled by $\exp[-\Lambda\mathrm{Tr}(\mathcal W^\dag_K \mathcal W_K)]$ with $\Lambda\to \infty$.
However, $K=2$ is anyway insufficient because $\mathcal W_2=0$ binds a single matrix up to a constant factor. Then instead of 4-fold open string degeneracy,
we get a much larger bunch of independent states for each $V$ between endpoints. Therefore we should take at least $K=3$.
The output space of $\hat{c}$ is spinorlike but only auxiliary.
The complete Hamiltonian traces $\hat{c}$ with $\hat{c}^\dag$ to get a scalar and reads
\begin{equation}
\hat{H}_w=\int D^KrD^Kv w(r,v)\mathrm{Tr}[\hat{c}^\dag(r,v)\hat{c}(r,v)]\label{haa}
\end{equation}
with some real function $w$ positive for the local link wiggling and configuration measure taken $K$ times.

The trace gives a scalar because of Pauli matrices multiplication $\sigma_j\sigma_k=\delta_{jk}I+i\epsilon_{jkl}\sigma_l$ for $jkl=1,2,3$ and
 $\mathrm{Tr}\sigma_k=0$ and so (\ref{haa}) is defined only in the string space with the spinor traced out to a scalar.

Before considering swap Hamiltonian note that already the space of ground states of wiggling is quite restricted. The only elementary operation -- swap
between fragments of different strings (or even the same), see Fig. \ref{wi2} -- applied twice must return to the original state. In other words, the double swap is identity and so there are
only two eigenspaces of the swap, with $\pm 1$ eigenvalue. Obviously, $+1$ would give a bosonic state while $-1$ is desired for fermions.
We can try to construct the swap annihilation operator like we did it for wiggling. Unfortunately, the swap counterpart of (\ref{slat}) is more complicated, having $16$ entries instead of $4$ matrix elements. We shall assume that the swap preserves the sum of nominal lengths of the swapping strings
but this is not obligatory.

In principle, we can simply generalize Slater matrix (\ref{slat}) and (\ref{slat2}) replacing $V$ with a tensor product of two links.
In addition, the tensor can be written in both representations, linking $A-B$ and $C-D$ or $A-D$ and $C-B$.
Let us denote such a tensor by a $2\times 2\times 2\times 2$ matrix $W$ with entries
$W_{abcd}$, $a,b,c,d=\pm$. For $A-B$ and $C-D$ links $V$ and $U$ respectively
we define $W_{abcd}=\bar V_{ab}\bar U_{cd}$ while for $A-D$ and $C-B$ links $V'$ and $U'$ respectively
we define $W_{abcd}=-\bar V'_{ad}\bar U'_{bc}$ (the $-$ sign is to get antisymmetric fermions, with $+$ we get bosons). Generalizing (\ref{slat}) we define
\begin{equation}
\mathcal S_K=\sum_\tau\mathrm{sgn}\tau \bar f^{\tau(1)}W^{\tau(1)}\bar f^{\tau(2)}W^{\tau(2)}\cdots \bar f^{\tau(K)}W^{\tau(K)}\label{slatw}
\end{equation}
where $W^j$ can be either of linkings with appropriate $f^j$
and
\begin{equation}
\hat{c}(r,v)=\sum_\tau\mathrm{sgn}\tau \bar f^{\tau(2)}W^{\tau(2)}\cdots \bar f^{\tau(K)}W^{\tau(K)} \langle rv^{\sigma(1)}|\label{slat2w}
\end{equation}
The difference from wiggling is that now there are maximally 16 linearly independent matrices $W$ so $\hat{c}$ vanishes for $K>17$
while for $K<17$ we need the constraint $\mathcal S_K=0$ by adding $\delta(\mathcal S_K)\sim \exp[-\Lambda\mathrm{Tr}(\mathcal S^\dag \mathcal S)]$. The optimal choice is $K=17$ with generic random set of linkings. If such high $K$ seems awkward we can 
take a lower value. The minimal $K=2$ would require $W^1_{abcd}=\bar V_{ab}\bar U_{cd}$ and $W^2_{abcd}=-\bar V'_{ad}\bar U'_{cb}$ but the constraint $\mathcal S_2=0$
results in the proportionality condition
\begin{equation}
\bar V_{ab}\bar U_{cd}\propto \bar V'_{ad}\bar U'_{cb}\label{vv1}
\end{equation}
where $V,U,V',U'$ are matrices of all links.  Unfortunately, it holds only if all the matrices are singular, see Appendix.
 Therefore (\ref{vv1}) and $\mathcal S_2=0$ will be only satisfied if all the involved matrices are singular (e.g. projection matrices, appearing in the asymptotic limit $|\mathrm{Im}\vec{v}|\to\infty$), which is the case we wanted to avoid. Even $\mathcal S_3=0$ only if some of the matrices are singular (assuming they 
 are not all for the same linking) and $\mathcal S_4=0$ if only one $W$ is from one of the linkings while three are from the other linking
(but already pairs for each linking can combine to the same projection), see Appendix.
Despite the above obstacles, we will explain that we can use even $K=2$ abandoning $\mathcal S_2=0$ constraint and construct annihilation operators based on (\ref{slat2w}),
\begin{eqnarray}
&&\hat{c}_{abcd}(v,u,v',u')= \\
&&\bar f(v)\bar f(u)\bar V_{ab}\bar U_{cd}\langle v' u'|+\bar f(v')\bar f(u')\bar V'_{ad}\bar U'_{cb}\langle v u|\nonumber
\end{eqnarray}
and
\begin{eqnarray}
&&\hat{H}_s=\int DvDv'DuDu'\sum_{abcd}\\
&&w(v,u,v',u')\hat{c}^\dag_{abcd}(v,u,v',u')\hat{c}_{abcd}(v,u,v',u')\nonumber
\end{eqnarray}
with some real function $w$ positive for a local swap.
Explicitly 
\begin{eqnarray}
&&\hat{H}_s=
\int DvDv'DuDu'w(v,u,v',u')
\label{sw3}\\
&&\left[|\bar f(v)\bar f(u)|^2|v'u'\rangle\langle v'u'|\mathrm{Tr}V^\dag V\mathrm{Tr}U^\dag U\right.\nonumber\\
&&
+|\bar f(v')\bar f(u')|^2|vu\rangle\langle vu|\mathrm{Tr}V^{\prime\dag}V'\mathrm{Tr}U^{\prime\dag}U'\nonumber\\
&&
+\bar f^\ast (v)\bar f^\ast(u)\bar f(v')\bar f(u')|v'u'\rangle\langle vu|\mathrm{Tr}V^\dag V' U^\dag U'\nonumber\\
&&
+\bar f^\ast (v')\bar f^\ast(u')\bar f(v)\bar f(u)|vu\rangle\langle v'u'|\left.\mathrm{Tr}V^{\prime\dag}V U^{\prime\dag}U\right]\nonumber
 \end{eqnarray}

 In contrast to wiggling, the state (\ref{stri}) is not an eigenstate of the above Hamiltonian with zero eigenvalue
 but it is not necessary. We can treat $\hat{H}_s$ as a small perturbation and check the average of $\hat{H}_s$ in the symmetric and antisymmetric state.
 It suffices to get a smaller average for the antisymmetric state which becomes stable in this way.
 Let us assume that the average length of the string/loop is much longer than the correlation length of $\vec{v}$. Then
 calculating the above-mentioned average we can assume a random spin state of the string endpoints because it will get randomized along the string.
If we extend $V$ to in the direction of endpoints $A$ and $B$, $U$ in $C$ and $D$, $V'$ in $A$ and $D$, and $U'$ in $C$ and $B$,
sufficiently far in such a way that $V$ and $V'$ have a long common matrix factor in $A$ direction, $U$ and $V'$ in $D$, $V$ and $U'$ i $B$
and $U$ and $U'$ in $C$ then the average of $|vu\rangle\langle v'u'|$ reads
 \begin{equation}
\pm \bar f^\ast(u)\bar f^\ast(v)\bar f(u')\bar f(v')\mathrm{Tr}V^\dag  V' U^\dag  U'
\end{equation}
 with $+$ for the symmetric and $-$ for the antisymmetric state, up to some positive prefactor.
 The last two lines of (\ref{sw3}) are equal
\begin{eqnarray}
&&\pm 2\sum_{v,u,v',u'}w(v,u,v',u')|g(r,v)\bar f (v)\bar f(u)\bar f(v')\bar f(u'))|^2\nonumber\\
&&\times|\mathrm{Tr}V^\dag V' U^\dag U'|^2\label{ferbos}
\end{eqnarray}
which is positive for the symmetric and negative for the antisymmetric (fermionic) state.
The antisymmetric state has then lower energy (in the first order) than the symmetric and so it is stable.

We have ignored string crossing. Like lines, the strings in 3D can cross each other at particular points and times.
In principle it could lead to some additional interaction, e.g. preventing from crossing by some repulsion or forcing a discontinuous crossing. 
We could modify wiggling or swapping by a factor controlling the relative position of strings but it will not change the general idea.
Assuming a small density of strings (defined as nominal length per volume) times the interstring interaction length (average nominal length the other string that a  given point of a string interacts with, scaled by Hamiltonian), the repulsion will be as negligible as e.g. collisions in an ideal gas.

The ambiguity or flexibility of the choice of swapping and wiggling terms cannot alter the bosonization, because the effective state
depends on the reduced number of degrees of freedom (endpoint position and spin), just like the phase in a quantum phase transition is described by an effective (order) parameter.

The Hamiltonian can have eigenstates whose energies approach zero e.g. by slowly varying wave functions. However, we can boost the prefactors $w$ to increase the relevant variation lengthscale beyond detectable infrared bound, just like long photons are irrelevant.


\begin{figure}
\includegraphics[scale=.4]{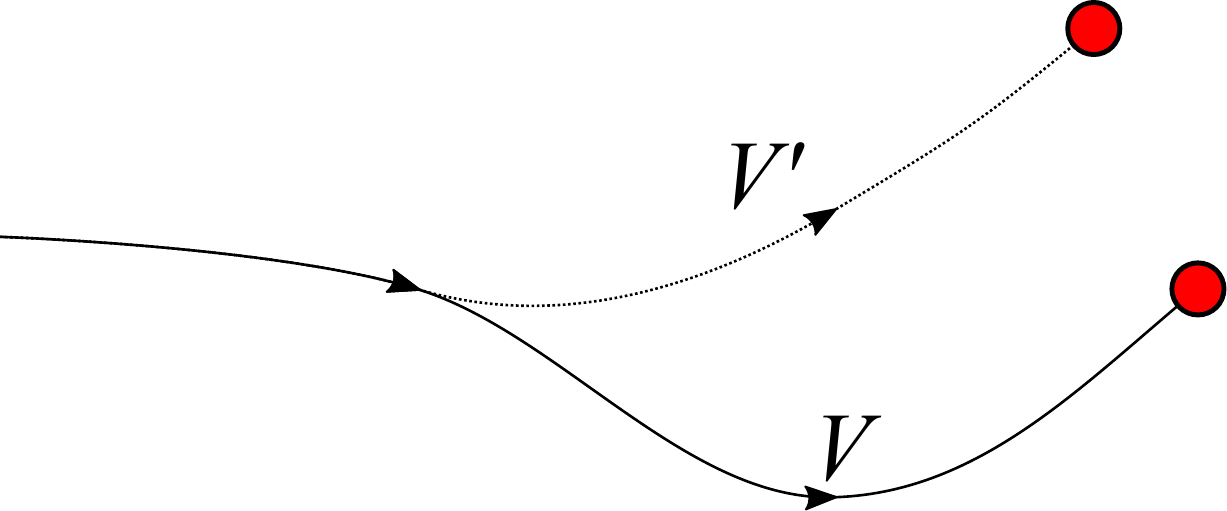}
\caption{Local wiggling of the string endpoint.}
\label{wi3}
\end{figure}

\begin{figure}
\includegraphics[scale=.4]{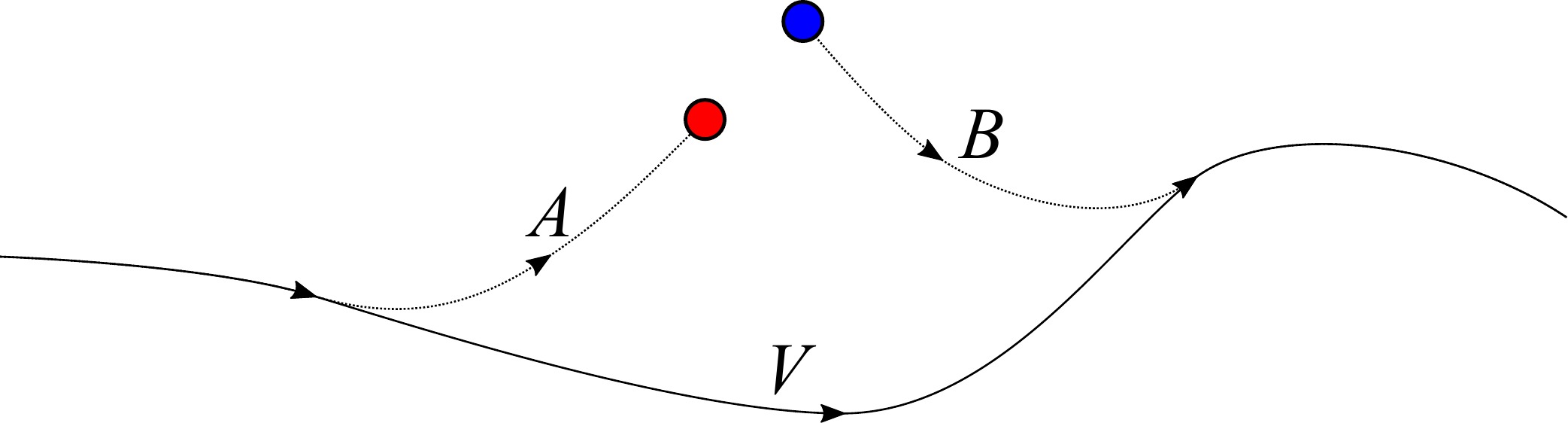}
\caption{Splitting of the string to create to endpoints $L$ and $R$.}
\label{mas}
\end{figure}

\section{Endpoints dynamics}

So far we have considered states with fixed endpoints. All these states are degenerate with zero energy, optionally (if using $\mathcal S_K$ with $K=2$ or generally $K<17$) corrected by a constant term $\langle \hat{H}_s\rangle$ assuming the swaps are rare.
Since the so far considered Hamiltonian has not changed positions of endpoints, the states of open strings are parametrized by this position leaving them degenerate. For instance, a state with a localized endpoint has still zero energy. Towards our final goal -- effective reconstruction of Dirac Hamiltonian,
we will need to keep minimum energy for delocalized states, such that $f$ depends only on relative positions, i.e.
$f(\vec{r}+\vec{r}_0,v)=f(\vec{r},v)$ for an arbitrary $\vec{r}_0$. This is possible by adding any positive term tracking dependence on $\vec{r}_L$ and $\vec{r}_R$,  applying the same wiggling term (\ref{haa}) as in the case of the internal part of the string, see Fig. \ref{wi3}. A possible Hamiltonian reads
\begin{equation}
\hat{H}_e=\alpha \sum_{P=R,L}\int DrDv |\bar{f}|^2\left(\frac{\delta |rv\rangle/\bar{f}^\ast}{\delta\vec{r}_P}\right)\cdot\left(\frac{\delta\langle rv|/\bar{f}}{\delta\vec{r}_P}\right)
\end{equation} 
with sufficiently large positive $\alpha>0$ and $f=\bar{f}g$ such that $g$ is independent of $\vec{r}_{L/R}$. 
Here the functional derivatives at the endpoints are taken in the one-sided limit along the string, i.e. $\delta/\delta\vec{r}_R=\lim_{s\to s_{R-}}\delta/\delta\vec{r}(s)$, assuming $f$ regular or regularized at $s_{R/L}$.
Then the only state with zero energy is the absolutely delocalized one. However, the states (\ref{multi}) slowly varying,
\begin{equation}
|\vec{k},a;\vec{q},b\rangle=\int \frac{d\vec{x}d\vec{y}}{\Omega}\label{kqe}\exp(i\vec{k}\cdot \vec{x}+i\vec{q}\cdot\vec{y})|\vec{x},a;\vec{y},b\rangle\nonumber
\end{equation}
 have the first order effective Hamiltonian
\begin{eqnarray}
&&\langle\vec{k'},a';\vec{q'},b'|\hat{H}_e|\vec{k},a;\vec{q},b\rangle=\nonumber\\
&&\alpha(|\vec{k}|^2+|\vec{q}|^2)\delta_{aa'}\delta_{bb'}\delta(\vec{k}-\vec{k'})\delta(q-\vec{q'})\label{regg}
\end{eqnarray}
 which goes to zero asymptotically for $k,q\to 0$. The above Hamiltonian generalizes immediately to $N$ open string.
In position space $\vec{k}=-i\nabla$ so $|\vec{k}|^2=-\Delta$.
Note that $\Delta$ term is absent in Dirac equation but we can make $\alpha$ so small to keep this term negligible in the accessible regime.

Now, suppose we add another very small term 
\begin{equation}
\hat{H}_P=\int DrDr'DvDv'\;h_P(\vec{r'}\vec{v'};
\vec{r}\vec{v})|r'v'\rangle\langle rv|
\end{equation}
where $h$ is rotationally invariant functional of $\vec{r}$, $\vec{v}$, $\vec{r'}$, $\vec{v'}$
Invariance essentially requires that $h$ depends on scalars (pseudocalars), i.e. scalar or mixed products
$\vec{r}\cdot\vec{r}$, $\vec{r}\cdot\vec{v}$, $\vec{r'}\cdot \vec{v'}$, $\vec{r}\cdot (\vec{r'}\times\vec{v})$ etc.
we also demand that $\langle \hat{H}_{L/R}\rangle=0$ for $k=q=0$ (reference state). From perturbation theory,
the first nonvanishing correction due to $\hat{H}_{L/R}$ to the effective Hamiltonian on the states (\ref{kqe}) is linear in $\vec{k}$
Since the states have already spinor structure
\begin{eqnarray}
&&\langle\vec{k'},a';\vec{q'},b'|\hat{H}_L+\hat{H}_R|\vec{k},a;\vec{q},b\rangle=\delta(\vec{k}-\vec{k'})\delta(\vec{q}-\vec{q'})\nonumber\\
&&[c_L(\vec{k}\cdot\vec{\sigma})_{aa'}\delta_{bb'}+c_R(\vec{q}\cdot\vec{\sigma})_{b'b}\delta_{aa'}+O(k^2+q^2)]\label{kkk}
\end{eqnarray}
An example reads $\hat{H}_P=$
\begin{equation}
\int DrDv \left(\left(i\xi_P\vec{v}^i_P+\omega_P\frac{\delta}{\delta\vec{v}^r_P}\right)|rv\rangle\right)\cdot\frac{\bar{f}\delta\langle rv|/\bar{f}}{\delta\vec{r}_P}+\mathrm{H.c.}
\end{equation} 
with $\vec{v}=\vec{v}^r+i\vec{v}^i$ (real and imaginary part).
For large $\vec{k}$ we will get nonlinearities and/or interaction (excitations are no longer independent).
This scale is determined by the density of string and interactions, but in our thermodynamic limit the linear, noninteracting regime of low $\vec{k}$
always exists.

The last term we need is string splitting, Fig. \ref{mas}, necessary to recover mass in Dirac equation. It will change the number of open strings but remains local.
The split Hamiltonian, connecting and disconnecting string, reads in general
\begin{eqnarray}
&&\hat{H}_m=\int DrDvDr'Dv' ds\\
&&h_m(r'v'(\to s),r'v'(s\to);rv)|r'v'\rangle\langle rv|+\mathrm{H.c.}\nonumber
\end{eqnarray}
where the configuration $r'v'$ is split into the left and right part (see Fig. \ref{mas}) preserving the total nominal length.
Here  $rv(\to s)$ denotes the part of string/loop $rv$ ending at $s$ while $rv(s\to)$ is the part starting at $s$ scanned along all strings/loops. 
The Hamiltonian must be local and rotationally invariant.
 The Hamiltonian is still local, i.e. it does not know if the string before splitting is open or closed.
 In the first case, the output is two open strings while in the second case the output is one open string.
 In any case the number of open strings increases (decreases) by one for splitting (joining).
The simplest example reads
\begin{equation}
\hat{H}_m=\int DrDv ds m|rv(\to s);rv(s\to)\rangle\langle rv|+\mathrm{H.c.}
 \end{equation}
 It essentially only breaks/joins the string/loop leaving the configuration unchanged.
 However, once the endpoints are created, the endpoint dynamics uncouples them. The value of mass $m$ must be certainly small within the validity range
 of linear approximation.

\section{Reconstructing Dirac equation}

 Now we want to lift this degeneracy and recover Dirac dynamics, $i\partial_t=\hat{H}$.
The evolution  of the excitation
\begin{eqnarray}
&&\phi|\Omega\rangle+\sum_{ab}\int d\vec{x}d\vec{y}\psi_{ba}(\vec{x},\vec{y})|\vec{x}_{La},\vec{y}_{Rb}\rangle+\\
&&\sum_{abcd}\int d\vec{x}d\vec{y}d\vec{z}d\vec{w}\xi_{dcba}(\vec{x},\vec{y},\vec{z},\vec{w})|\vec{x}_{La}\vec{z}_{Lc}\vec{y}_{Rb}\vec{w}_{Rd}\rangle\nonumber
\end{eqnarray}
with $\xi_{dcba}(\vec{x},\vec{y},\vec{z},\vec{w})=-\xi_{bcda}(\vec{x},\vec{w},\vec{z},\vec{y})=-\xi_{dabc}(\vec{w},\vec{y},\vec{z},\vec{x})$
reads
\begin{eqnarray}
&&i\partial_t\phi=-m\sum_a\int d\vec{x} \psi_{aa}(\vec{x},\vec{x}),\:i\partial_t\psi_{ba}=\label{dir}\\
&&(i\nabla_x-\vec{A}(\vec{x}))\cdot(\psi\vec{\sigma})_{ba}-(i\nabla_y+\vec{A}(\vec{y}))\cdot(\vec{\sigma}\psi)_{ba}
\nonumber\\
&&-(A_0(\vec{x})-A_0(\vec{y}))\psi_{ba}-m\phi\delta_{ba}\delta(\vec{x}-\vec{y})\nonumber\\
&&-m\int d\vec{z} \sum_c\xi_{ccba}(\vec{x},\vec{y},\vec{z},\vec{z})\nonumber
\end{eqnarray}
The mass term changes the number of excitation pairs. It is easy but lengthy to write down evolution for higher excitations.
Without the mass, the evolution is simply an analog of the equation for $\psi$ while the mass $m$ allows jumps between one more or one less
pair.
To recover (\ref{dir}) without gauge potential we simply need to have $c_R=1=-c_L$ and negligible $\alpha$ in (\ref{kkk}) and (\ref{regg}), respectively.

The $\Delta$-term  is critical to keep the finite bottom of the Fermi sea.
The effective Dirac Hamiltonian we reconstructed will have its ground state different from $|\Omega\rangle$ because filling the negative energy levels
will lower the total energy. Without the $\Delta$-term the levels would continue until cutting all strings into short intervals, ruining the model.
Remember that the sign of energy of levels far from zero does not depend on chirality ($L/R$) but helicity (sign of eigenvalue of $\vec{k}\cdot\vec{\sigma}$).
To prevent such a collapse, at very large $|\vec{k}|$ the energy must go up so that further cutting the strings becomes energetically unfavorable.
The energy scale can be set safely far from the expected regime of validity of Dirac equation. For large $\vec{k}$ the fermions may be also  no longer
noninteracting. The $\Delta$-term can be viewed as an analog of fermion doubling \cite{doub1,doub2,doub3}, occurring when discretizing space.
The energy crosses zero at some large value of $\vec{k}$ which could be identified as an extra quasiparticle but such an excitation is unlikely
because of momentum conservation (e.g. a background field Fourier component of the comparable $\vec{k}$). This quasiparticle will be important
in renormalization when dynamics of field is included, but it is beyond the scope of this work.

\begin{figure}
\includegraphics[scale=.5]{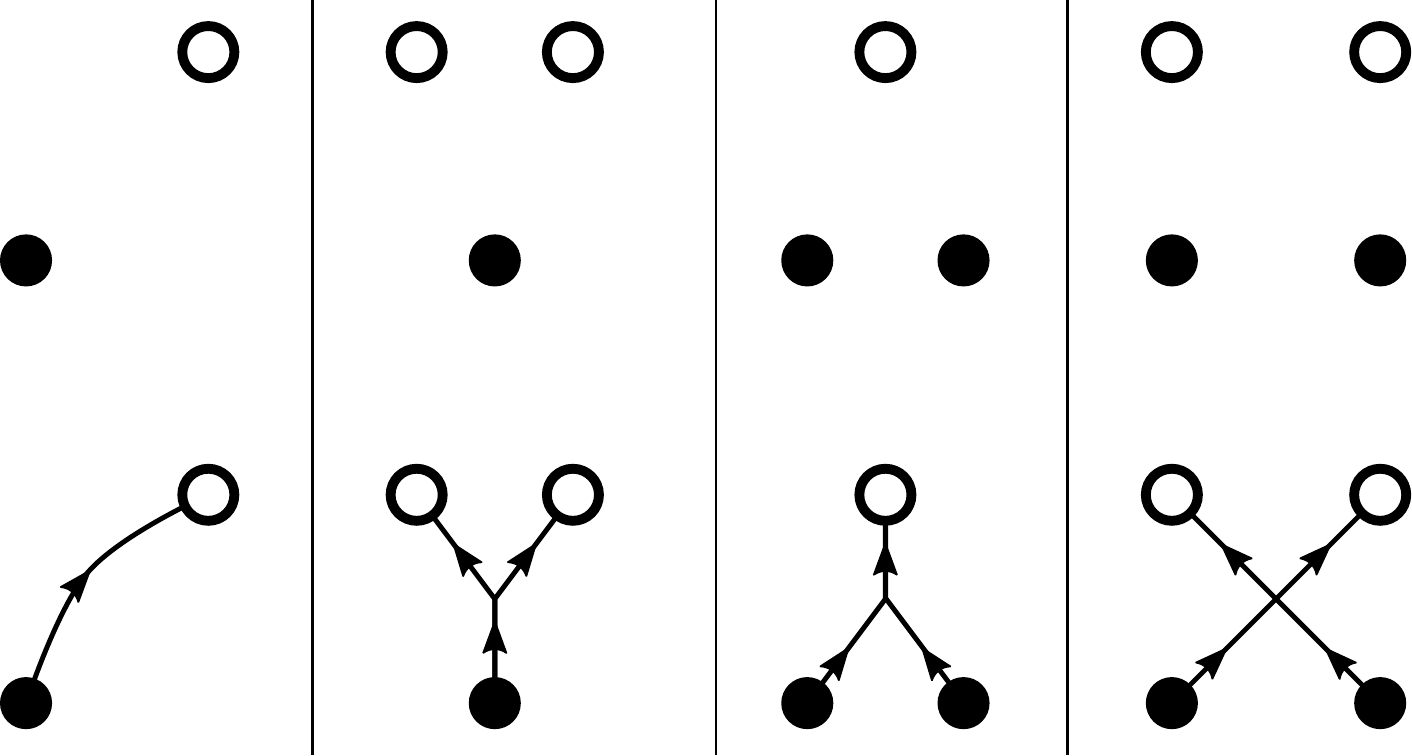}
\caption{Comparison between hopping and moving of point particles.
Hopping (upper row) depends only on the initial (black) and final (white) position.
Moving (lower row) depends on the continuous path from the initial to the final position.
A complicated moving contains splitting or hub points.}
\label{hopp}
\end{figure}

\begin{figure}
\includegraphics[scale=.4]{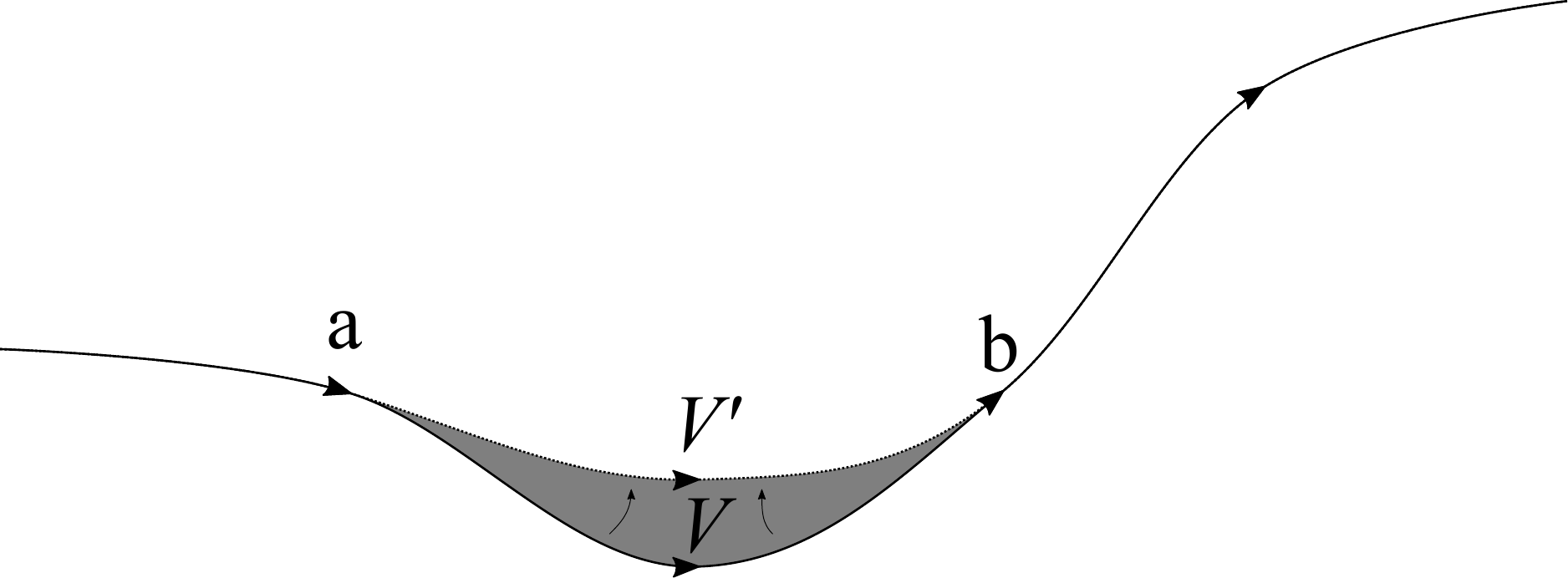}
\includegraphics[scale=.4]{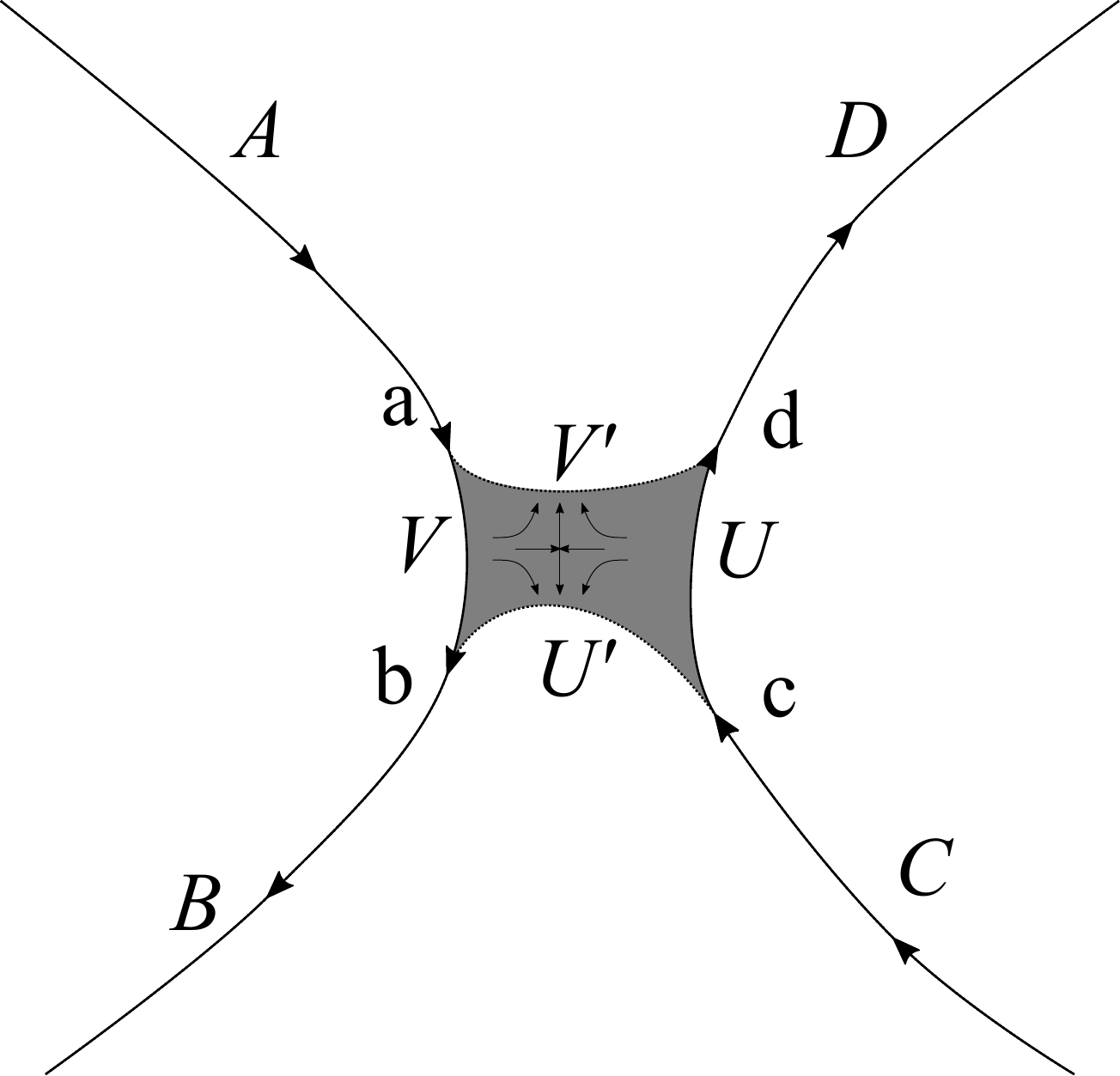}
\includegraphics[scale=.4]{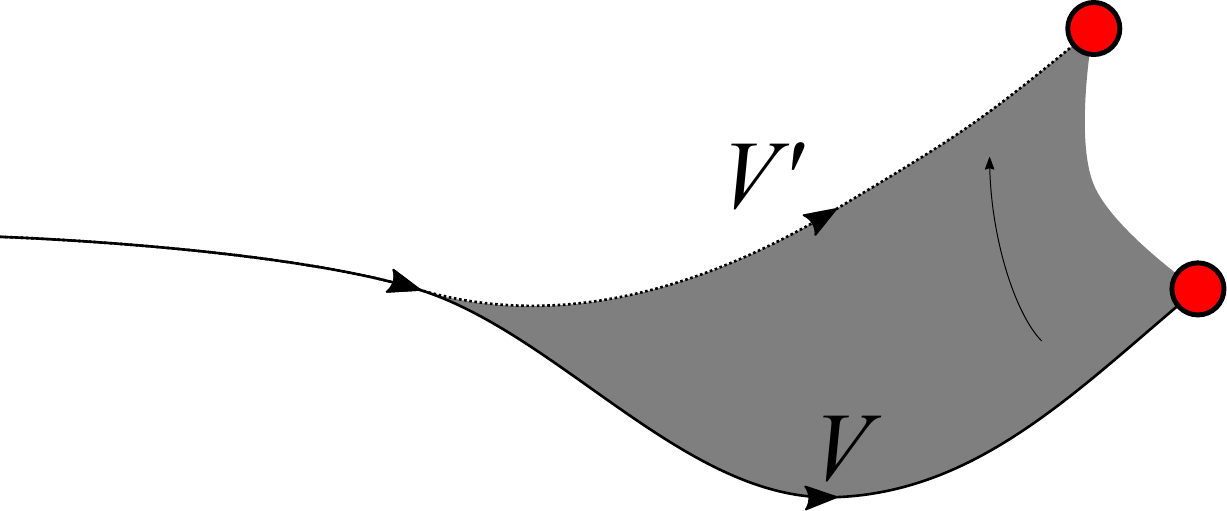}
\includegraphics[scale=.4]{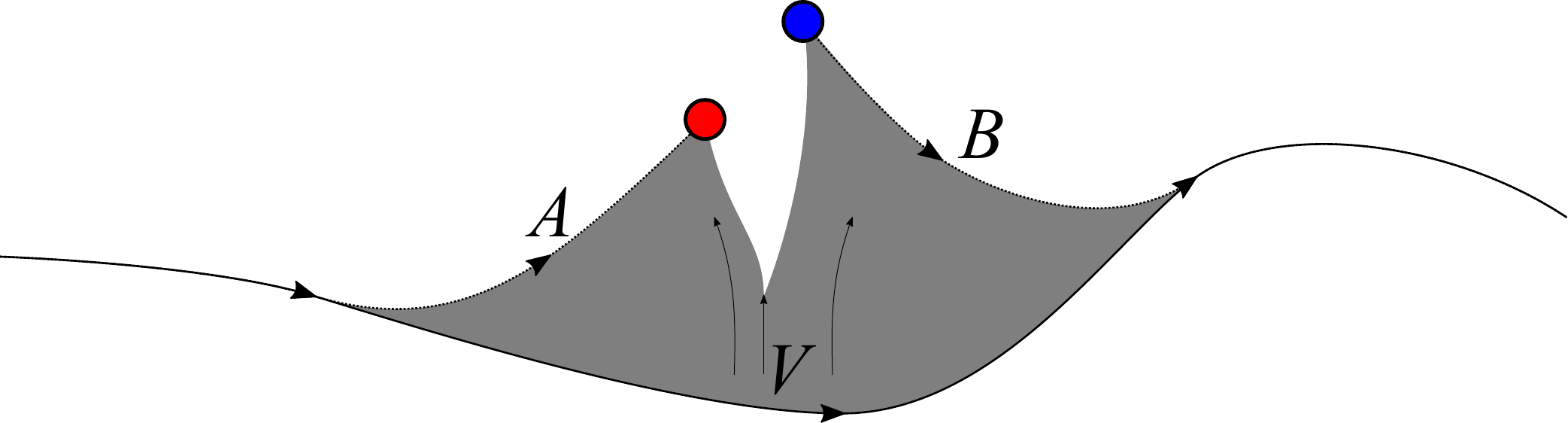}
\caption{Dragging counterparts of Figs. \ref{wi1}, \ref{wi2}, \ref{wi3}, \ref{mas}.
In the case of swapping there is a hub/saddle of tangent vectors $\partial_\lambda\vec{r}$.
The endpoint dragging will produce edge path. The mass term contains a splitting/tearing point.}
\label{drag}
\end{figure}

To incorporate the influence of the gauge potential we could of course simply add appropriate potentials to endpoint dynamics.
Instead, we propose a construction which not only recovers (\ref{dir}) but requires only electromagnetic fields (not potentials)
in the Hamiltonian. We modify $f$ in the definition of the string wave function,
\begin{equation}
\tilde f(r,v)=f(r,v)\exp\int ds \vec{A}(\vec{r}(s))\cdot d\vec{r}/ids
\end{equation}
 Let us consider the gauge covariant derivatives
\begin{eqnarray}
&&\tilde\delta_{\vec{r}(s)}\langle rv|=
\frac{\delta\langle rv|}{\delta\vec{r}(s)}-i\vec{B}(\vec{r})\times 
\frac{d\vec{r}}{ds}\langle rv|\\
&&\tilde\delta_{\vec{r}(s)}|rv\rangle=
\frac{\delta| rv\rangle}{\delta\vec{r}(s)}+i\vec{B}(\vec{r})\times 
\frac{d\vec{r}}{ds}|rv\rangle\nonumber
\end{eqnarray}
with $\vec{B}=\nabla\times\vec{A}$ and $\vec{E}=-\partial_t\vec{A}-\nabla A_0$,
and gauge drag
\begin{equation}
\tilde h(r'v';rv)=h(r'v';rv)\exp\int ids d\lambda (\vec{B}(\vec{r})\times\partial_s\vec{r})\cdot\partial_\lambda\vec{r},
\end{equation}
resembling Kogut-Susskind plaquette Hamiltonian \cite{kogut}
with $\vec{r}'=\vec{r}(\lambda_1)$ and $\vec{r}=\vec{r}(\lambda_0)$ with $\vec{r}(s,\lambda)$ spanning the surface between $\vec{r}$ and $\vec{r'}$
where they differ.
It essentially means that the Hamiltonian connecting two different trajectories depends also on the path (drag along a sheet) between them.
The situation is analogous to pointlike particles with hopping. The hopping means that we care only about the initial and final point.
Replacing hopping by \emph{moving} we keep track of the continuous path between the points, see Fig. \ref{hopp}.
In our case, the wiggling/swapping containing the information only about the initial and final trajectory will be replaced by the \emph{dragging} when we scan the whole two-dimensional sheet between the trajectories. In some cases, the dragging -- like moving -- contains splitting or hub points, see Fig. \ref{drag}.


Now we recover (\ref{dir}) by replacing $f$ with $\tilde{f}$ in (\ref{stri}) while
modifying $\hat{H}$ by replacing kernel $h$ in (\ref{kerne}) by $\tilde{h}$ and $\delta_{\vec{r}}$ by $\tilde{\delta}_{\vec{r}}$ in derivative-based kernels
and adding $\hat{H}_E$,
\begin{equation}
\hat{H}\to\hat{\tilde{H}}+\hat{H}_E
\end{equation}
with the \emph{electric} term
\begin{equation}
\hat{H}_E=-\int DrDv  ds\vec{E}(\vec{r})\cdot \frac{d\vec{r}}{ds}|rv\rangle\langle rv|
\end{equation}
All fields here depend also on time, hidden in the notation for brevity.

\section{Discussion}

We have proposed an improved string-net model of bosonization of fermions recovering Dirac dynamics in low-energy regime.
It is based on $SU(2)$ Wilson lines along strings connecting opposite charges of loops. We postulated the family of ground states
and deliberately defined the Hamiltonian such that these states have the lowest energy zero, with help of Slater determinant.
The final reconstruction of Dirac dynamics required some constraints on the perturbative part. The model is rotationally invariant,
bosonic, spin $1/2$ appears only effectively and potentials have been replaced by fields. It is spatially local but obviously we lost Lorentz invariance.

Despite the minimal goal achieved, there are many puzzles arising in this concept demanding further research. For instance, the general effective Dirac-like equation we could obtain is
\begin{equation}
i\partial_t
\begin{pmatrix}
\psi_L\\
\psi_R
\end{pmatrix}=i\vec{\sigma}\cdot\nabla
\begin{pmatrix}
c_L\psi_L\\
c_R\psi_R
\end{pmatrix}
+m\begin{pmatrix}
\psi_R\\
\psi_L
\end{pmatrix}-\alpha\Delta\begin{pmatrix}
\psi_L\\
\psi_R
\end{pmatrix}\label{dirv}
\end{equation}
We have at present no clue why the free parameters $c_L$, $c_R$ and $\alpha$ satisfy $c_L+c_R=0$ (then we can rescale time to $c_R=1$) and small $\alpha$
but it is expected to be connected with restoring effective Lorentz invariance. 
Another task is to include the dynamics of fields $\vec{E}$ and $\vec{B}$, while here they are only background. 
They can appear among other excitations beyond the Dirac fermions (e.g. controlled by the magnetic 
flux traversed by the string) but one has the renormalization to deal with. It is also worth to generalize the model beyond electrodynamics and try to include Lorentz symmetry (e.g. by adding extra dimensions) or prove that it is impossible.

\begin{figure}
\includegraphics[scale=.5]{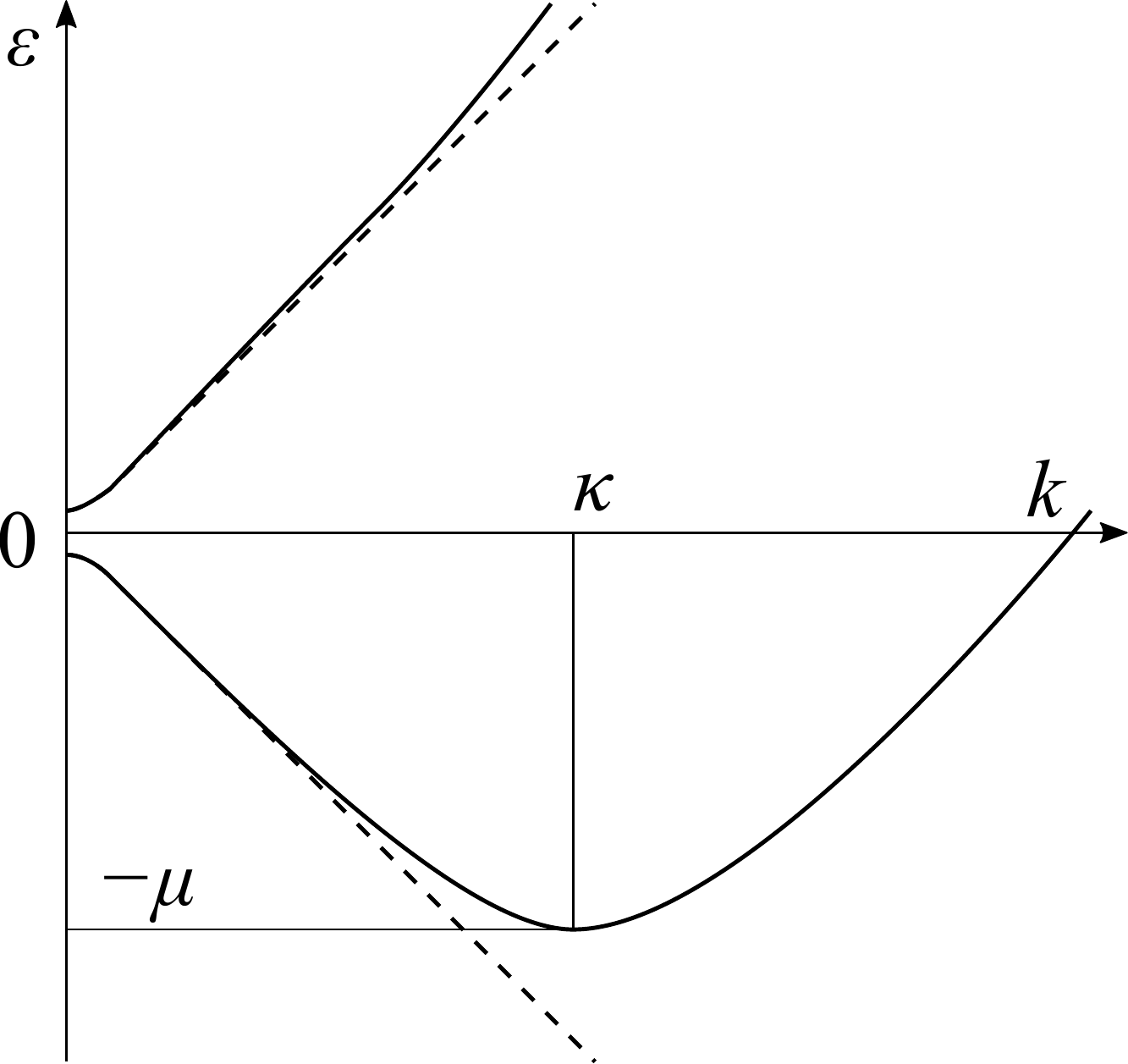}
\caption{The difference between Dirac (dashed) and our (solid) dispersion relation $\epsilon(k)$.
In our case there is an absolute minimum at $k=\kappa$, $\epsilon=-\mu$}
\label{ekk}
\end{figure}

The $\Delta$-term  can lead to a quantum phase transition.
The eigenvalue of $i\partial_t=\epsilon$ and $-i\nabla=\vec{k}$, is depicted in Fig. \ref{ekk}.
The fermion antisymmetry implies the ground state with the single excitation with $\epsilon<0$ occupied only once -- the Fermi sea.
In the original Dirac dispersion, $\epsilon^2=k^2+m^2$ is unbounded from below leading to the breakdown of the string-net into short pieces.
Adding our $\Delta$-term we get a minimum at $k=\kappa$, $\epsilon=-\mu$. If $\kappa^{-1}$ is much larger than the string correlation distance then our linear and independent approximation (no higher powers of $k$, no coupling between levels) is valid
in the whole Fermi sea. However, if the swap Hamiltonian like (\ref{ferbos}) is small (e.g. for a small density of strings) then the energy difference
between antisymmetric fermions and symmetric bosons competes with the Bose-Einstein condensation at $\epsilon=-\mu$ (bosons, unlike fermions, will simply
occupy the same lowest state). Both states are string-nets but their properties are fundamentally different.
It is an open question how to model best this transition.

Summarizing, the presented model is only an intermediate step toward the full bosonization of fermions, but it shows that -- sacrificing relativity
-- some construction exists.  The deviations from perfect Dirac equation (\ref{dirv}) can be experimentally tested but
due to corrections from theories beyond electrodynamics the clearest signature at this stage would be a violation of Lorentz invariance.
It is also possible that further exploration of excited states will allow us to identify other known or new emergent particles.

\section*{Acknowledgement}
I thank P. Jakubczyk and P. Chankowski for inspiring discussion about the subject.

\section*{Appendix Swap Condition for $K\leq 4$} 

 We will show that swap condition $\mathcal S_K=0$ is impossible for $K=2,3$ and nonsingular links, $K=4$ and nonsingular links if on of linkings is represented only once.
Case $K=2$.
If $V$ and $U$ are invertible, then
\begin{equation}
\delta_{ab}\delta_{cd}\propto (V^{-1}V')_{ad}(U^{-1}U')_{cb}
\end{equation}
If some $(V^{-1}V')_{ad}$ element is nonzero then only $(U^{-1}U')_{da}$ is nonzero so $U^{-1}U'$ has zero determinant and $U'$ cannot be invertible
and similarly $V'$.
Since matrices in (\ref{vv1}) must have equal ranks and $\mathrm{rk}V\otimes U=\mathrm{rk}V\mathrm{rk}U=\mathrm{rk}V'\mathrm{rk}U'=1$
contradicts the assumption that $V$ and $U$ are invertible. If $V$ is invertible then rank implies that either $V'$ or $U'$ is invertible, too.
If both $V$ and $U'$ are invertible then $(U^{\prime -1}U)_{cd}\delta_{ab}\propto (V^{-1}V')_{ad}\delta_{cb}$.
Taking $a=b\neq c$ we see that $U^{\prime -1}U$ vanishes, contradiction.

Case $K=3$.
We will show that linear dependence of $W^1_{abcd}=V^1_{ab}U^1_{cd}$, $W^2_{abcd}=V^2_{ab}U^2_{cd}$
and $W^3_{abcd}=V'_{ad}U'_{cb}$ implies singularity of at least one of matrices $V^1$, $V^2$, $U^1$, $U^2$, $V'$, $U'$.
Suppose all they are nonsingular. By scaling, we get $W^3=W^1+W^2$, giving 16 equations
\begin{equation}
V^1_{ab}U^1_{cd}+V^2_{ab}U^2_{cd}=V'_{ad}U'_{cb}
\end{equation}
We multiply the above set of equations
by $(V^1)^{-1}_{\alpha a}(U^2)^{-1}_{d\gamma}$ summing over $a$ and $d$ and replacing $\alpha$ and $\gamma$ back to $a$ and $d$ respectively to get
\begin{equation}
\delta_{ab}\tilde{U}^1_{cd}+\tilde{V}^2_{ab}\delta_{cd}=\tilde{V}'_{ad}U'_{cb}
\end{equation}
with $\tilde{U}^1=U^1(U^2)^{-1}$, $\tilde{V}^2=(V^1)^{-1}V^2$, $\tilde{V}'=(V^1)^{-1}V'(U^2)^{-1}$
Now multiply the result by $(\tilde{V}')^{-1}_{\alpha a}\tilde{V}'_{b\beta}$  and sum over $a$ and $b$ replacing finally $\alpha$ and $\beta$ by $a$ and $b$, respectively, to get
\begin{equation}
\delta_{ab}A_{cd}+B_{ab}\delta_{cd}=\delta_{ad}C_{cb}
\end{equation}
with $A=\tilde{U}^1$, $B=(\tilde{V}')^{-1}\tilde{V}^2\tilde{V}'$, $C=U'\tilde{V}'$.
Now for $abcd=-++-,+--+$ we get $C_{++}=C_{--}=0$, for $++-+,-+--,++--$ we get $C_{-+}=A_{-+}=B_{-+}=A_{--}+B_{++}$,
for $+-++,--+-,--++$ we get $C_{+-}=B_{+-}=A_{+-}=A_{++}+B_{--}$. For $+++-,---+,-+++,+---$ we get $A_{+-}=A_{-+}=B_{-+}=B_{+-}=0$
and for $++++,----$ we get $A_{++}+B_{++}=A_{--}+B_{--}=0$. The result is $C=0$ and $A=-B=\lambda I$, contradiction.

Case $K=4$.
The singularity is implied also in the case $K=4$ if
$W^1_{abcd}=V^1_{ab}U^1_{cd}$, $W^2_{abcd}=V^2_{ab}U^2_{cd}$
 $W^3_{abcd}=V^3_{ab}U^3_{cd}$, $W^4=V'_{ad}U'_{cb}$.
 As above we assume that all matrices are nonsingular, by the same multiplication the equation $W^4=W^1+W^2+W^3$
 can be simplified
 to
 \begin{equation}
\delta_{ab}A_{cd}+B_{ab}\delta_{cd}+ D_{ab}E_{cd}=\delta_{ad}C_{cb}
\end{equation}
with nondegenerate $A,B,C,D,E$.
For $abcd=+-+-,-+-+$ we get $D_{+-}E_{+-}=D_{-+}E_{-+}=0$ so one of each pair $(D_{+-},E_{+-})$ and $(D_{-+},E_{-+})$ must vanish.
Without loss of generality $D_{+-}=0$. Then for $+-++,+--+,+---$ we get $B_{+-}=C_{+-}$, $C_{--}=0$, $B_{+-}=0=C_{+-}$.
Moreover, if $D_{-+}=0$, too, then analogously $C_{++}=C_{-+}=0$ so $C=0$.
If $D_{-+},E_{+-}\neq 0$ and $E_{-+}=0$ then for $++-+,---+$ we get $A_{-+}=C_{-+}$, $A_{-+}=0=C_{-+}$.
Taking $-++-,+++-,-+++$ we get $D_{-+}E_{+-}=C_{++}$, $A_{+-}=-D_{++}E_{+-}$, $B_{-+}=-D_{-+}E_{++}$.
Taking $++--,--++,-+--,--+-$ we get $A_{--}+B_{++}=-D_{++}E_{--}$, $A_{++}+B_{--}=-D_{--}E_{++}$, $B_{-+}=-D_{-+}E_{--}$, $A_{+-}=-D_{--}E_{+-}$
so $E_{++}=E_{--}$ and $D_{++}=D_{--}$. Finally $++++,----$ give $A_{++}+B_{++}+D_{++}E_{++}=C_{++}$ and $A_{--}+B_{--}+D_{--}E_{--}=0$
so $A_{++}=A_{--}$, $B_{++}=B_{--}$ and  $C_{++}=0$ so again $C=0$, contradiction.

\end{document}